\documentclass[aps,prd,twocolumn,floatfix,longbibliography,nofootinbib]{revtex4-1}
\usepackage{graphicx}
\usepackage{amsmath,amssymb,siunitx,physics,mathtools}
\usepackage{color}
\usepackage{hyperref}

\newcommand{\apjl}{Astrophys. J. Lett.}

\begin{document}
\title{Quasinormal modes of Schwarzschild black holes on the real axis}
\author{Koutarou Kyutoku$^{1,2,3}$, Hayato Motohashi$^4$, Takahiro
Tanaka$^{1,2}$}
\affiliation{$^1$Department of Physics, Kyoto University, Kyoto
606-8502, Japan}
\affiliation{$^2$Center for Gravitational Physics and Quantum
Information, Yukawa Institute for Theoretical Physics, Kyoto University,
Kyoto 606-8502, Japan}
\affiliation{$^3$Interdisciplinary Theoretical and Mathematical Sciences
Program (iTHEMS), RIKEN, Wako, Saitama 351-0198, Japan}
\affiliation{$^4$Division of Liberal Arts, Kogakuin University, 2665-1
Nakano-machi, Hachioji, Tokyo 192-0015, Japan}

\date{\today}

\begin{abstract}
 We study the scattering of gravitational waves by a Schwarzschild black
 hole and its perturbed siblings to investigate influences of proposed
 spectral instability of quasinormal modes on the ringdown signal. Our
 results indicate that information of dominant ringdown signals, which
 are ascribed to the fundamental (i.e., least damping) quasinormal mode
 of unperturbed Schwarzschild black holes, is imprinted in the phase
 shift defined from the transmission amplitude ($1/A_\mathrm{in}$ in our
 notation). This approximately parallels the fact that the resonance of
 quantum systems is imprinted in the phase shift of the $S$-matrix. The
 phase shift around the oscillation frequency of the fundamental mode is
 modified only perturbatively even if the quasinormal-mode spectrum is
 destabilized by a perturbative bump at a distant location, signifying
 the stability of the ringdown signal. At the same time, the phase shift
 at low frequencies is modulated substantially reflecting the late-time
 excitation of echo signals associated with the quasinormal-mode
 spectrum after destabilization.
\end{abstract}

\maketitle

\section{Introduction} \label{sec:intro}

Quasinormal modes are believed to play a central role in clarifying
various properties of black-hole spacetimes (see, e.g.,
Refs.~\cite{2009CQGra..26p3001B,2011RvMP...83..793K} for
reviews). Physically, quasinormal modes are exponentially-decaying
monochromatic waves escaping to null infinity and to the event
horizon. They are excited in a universal manner as a response to
impinging fields
\cite{1970Natur.227..936V,1971ApJ...170L.105P,1972ApJ...172L..95G}
and/or the motion of matter
\cite{1971PhRvL..27.1466D,1972PhRvD...5.2932D,1978ApJ...224..643C,1979ApJ...230..870C},
and associated observable signals are called the ringdown
signals. Individual modes are characterized by a complex frequency,
i.e., the oscillation frequency and the decay width (or damping
time). Because they are determined by the mass and the spin for Kerr
(including Schwarzschild) black holes, analysis of the ringdown signals
tells us these two parameters
\cite{1989PhRvD..40.3194E,1992PhRvD..46.5236F}. Remarkably, because the
mass and the spin completely determine all the quasinormal modes,
consistency assessment of estimated parameters for individual modes will
enable us to test whether astrophysical black holes are really
represented by Kerr black holes in general relativity
\cite{2004CQGra..21..787D,2006PhRvD..73f4030B}. Furthermore,
quasinormal-mode spectra are expected to provide us with a hint to
quantum gravity
\cite{1983PhRvD..28.2929Y,1998PhRvL..81.4293H,2003PhRvL..90h1301D}.

In the last few years, it has been argued that the quasinormal-mode
spectrum could be destabilized by tiny perturbations to the potential of
various types
\cite{2021PhRvX..11c1003J,2022PhRvL.128k1103C,2022PhRvL.128u1102J}. In
particular, even the fundamental mode is destabilized when the potential
is modified at a large distance from the black hole
\cite{2022PhRvL.128k1103C}, dating back to
Ref.~\cite{1996PhRvD..53.4397N} (see also
Refs.~\cite{2020PhRvD.101j4009D,2021PhRvD.103b4019Q}). In realistic
situations, excitation of quasinormal modes is inevitably accompanied by
nontrivial perturbations due to the dynamical formation process of a
deformed black hole and/or surrounding material. Thus, the spectral
instability of quasinormal modes could be fatal to the program of
black-hole spectroscopy.

However, the spectral instability of quasinormal-mode spectra does not
destabilize observable ringdown signals, particularly on the damping
time scale of unperturbed quasinormal modes
\cite{2014PhRvD..89j4059B,2022PhRvD.106h4011B}. Rather, a lot of
evidence is accumulating that the ringdown signal is fairly universal
even for nonlinear perturbations. For example, various
numerical-relativity simulations of compact binary coalescences, both
for vacuum and nonvacuum systems, have clearly witnessed excitation of
unperturbed quasinormal modes in dynamical processes (see, e.g.,
Refs.~\cite{2019RPPh...82a6902D,2019PrPNP.10903714B}, for reviews). This
prediction is confirmed to large extent by a lot of binary-black-hole
observations \cite{2021arXiv211103606T}. These facts strongly suggest
that the ringdown signals from a perturbed black-hole spacetime do not
deviate nonperturbatively from those from the unperturbed spacetime,
even if the quasinormal-mode spectrum is destabilized in a catastrophic
manner. Stated differently, observable ringdown signals may not
necessarily be dictated by quasinormal-mode spectra (see also
Ref.~\cite{2022CQGra..39u7002J}), even though they actually are for
unperturbed Schwarzschild and Kerr spacetimes. For example, fitting of
observed ringdown signals against damped sinusoids likely returns
complex frequencies not included in the destabilized quasinormal-mode
spectra as the strongest component \cite{2022PhRvD.106h4011B}.

A probable reason for this irrelevance is that the quasinormal-mode
spectrum is defined merely as a result of analytic continuation of the
relevant Green's function in terms of the frequency. Mathematically,
quasinormal modes are defined as the poles on the complex frequency
plane of the Green's function satisfying outgoing (at null infinity) and
downgoing (at the event horizon) boundary conditions. Meanwhile, the
realistic signal should be decomposed into real-frequency modes by the
Fourier transformation. Thus, it is natural to expect that information
governing the ringdown signals, which are likely to be tightly related
to the poles in the unperturbed black-hole spacetime, is imprinted also
in the real-frequency Green's function (see also
Ref.~\cite{2017PhRvD..96h4002M} for real-frequency analysis of ringdown
and echo signals from exotic compact objects). In particular, the
fundamental mode, i.e., the quasinormal-mode pole closest to the real
axis, should have a decisive impact.

\begin{figure*}[tbp]
 \includegraphics[width=.95\linewidth]{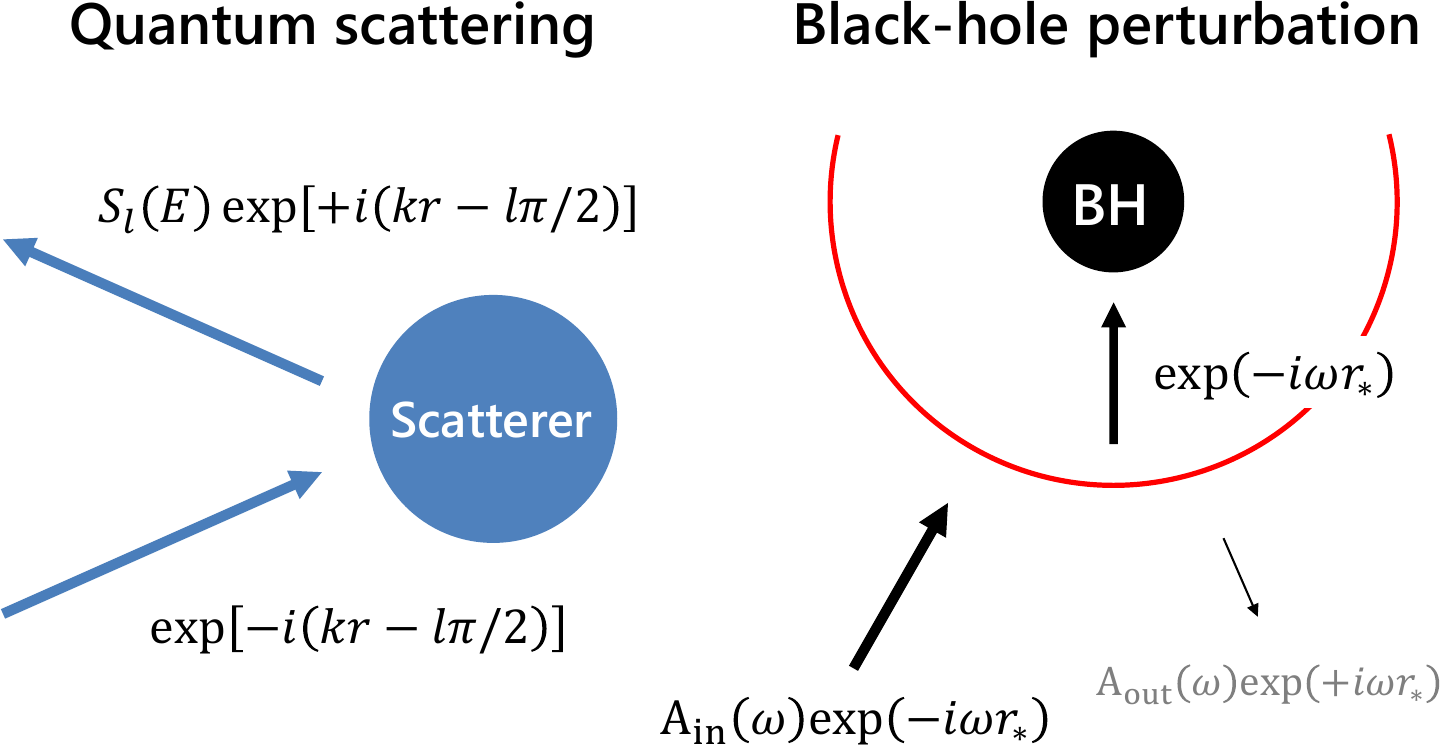} \caption{Schematic
 figure showing the correspondence between the quantum scattering and
 the black-hole perturbation. In the quantum scattering, we usually
 consider incident waves with the unit amplitude and scattered waves
 with the (complex) amplitude $S_l (E)$, and the poles of $S_l (E)$ give
 us information about the resonance (see, e.g., Ref.~\cite{taylor}). In
 this study on the black-hole perturbation, we consider ingoing waves
 with the (complex) amplitude $A_\mathrm{in} (\omega)$ and
 downgoing-wave components with the unit amplitude, and the zeroes of
 $A_\mathrm{in} (\omega)$ or equivalently the poles of $1/A_\mathrm{in}
 (\omega)$ actually give us information about the quasinormal modes. The
 red circle in the right panel is drawn to imply the barrier associated
 with the Regge-Wheeler potential \cite{1957PhRv..108.1063R}.}
 \label{fig:schematic}
\end{figure*}

In this study, we investigate the scattering of gravitational waves in
black-hole spacetimes to demonstrate that distant perturbations to the
potential hardly destabilize observable ringdown signals mainly
characterized by the unperturbed fundamental mode as anticipated
above. As an alternative to the poles on the complex frequency plane, we
propose that the phase shift defined from the transmission amplitude of
the ingoing waves enables us to infer the properties of unperturbed
quasinormal modes in an approximate but stable manner. This tool is
taken from the resonance in quantum systems, and the correspondence
assumed here is sketched in Fig.~\ref{fig:schematic} (see
Sec.~\ref{sec:time_freq} for detailed discussions). In
Sec.~\ref{sec:time}, by solving the time evolution of a Gaussian wave
packet, we check that the ringdown signal is stable even under
perturbations that destabilize the quasinormal-mode frequency of the
fundamental mode. Then, we explain our method to extract useful
information from the real-frequency scattering problem. By analyzing a
toy model in Sec.~\ref{sec:box}, we first explain how the spectral
instability is relevant only for complex frequencies required by
analytic continuation and next verify the utility of the phase shift as
a real-frequency indicator of dominant modes in the ringdown
signal. Equipped with this tool, we return to Schwarzschild black holes
and their perturbations in Sec.~\ref{sec:rw}. Section \ref{sec:summary}
is devoted to a summary and discussions.

Throughout this paper, we adopt the geometric unit in which $G=c=1$,
where $G$ and $c$ are the gravitational constant and the speed of light,
respectively. The mass of the black hole is denoted by $M$. The Fourier
transformation is performed with the convention that
\begin{equation}
 f(t) = \int_{-\infty}^\infty \tilde{f} ( \omega ) e^{-\mathrm{i} \omega
  t} \dd{\omega} .
\end{equation}
The complex quasinormal-mode frequency is decomposed into the
oscillation frequency $\omega_\mathrm{R}$ and the decay width $\Gamma$
by $\omega = \omega_\mathrm{R} - \mathrm{i} \Gamma /2$.

\section{Stability of ringdown signals under spectral instability}
\label{sec:time}

To set the stage for this study, we first show that the ringdown signals
in the Schwarzschild spacetime associated with the scattering of a
Gaussian wave packet is stable under perturbations of the effective
potential even if the fundamental mode is destabilized
\cite{2022PhRvL.128k1103C} (but see also Ref.~\cite{2022CQGra..39k5010G}
for possible alternative definitions of stability in terms of different
norms). The significant difference is observed only as late-time
excitation of echoes between the primary potential barrier and the
perturbation. A similar analysis has recently been made in
Ref.~\cite{2022PhRvD.106h4011B} for slightly different perturbations,
and our results seem consistent with theirs qualitatively. This analysis
motivates us to search for the signature of unperturbed quasinormal
modes, particularly the fundamental mode, from real-frequency solutions
of the scattering problem.

\subsection{Setup} \label{sec:time_setup}

Time evolution of linear waves represented by $\phi$ around a black hole
is governed by hyperbolic equations with effective
potentials. Specifically, after spherical harmonic decomposition (whose
indices are suppressed here for ease of notation), governing equations
reduce to (see, e.g., Ref.~\cite{1983mtbh.book.....C} for comprehensive
reviews)
\begin{equation}
 \pdv[2]{\phi}{t} = \pdv[2]{\phi}{r_*} - V (r_*) \phi (t,r_*) ,
  \label{eq:wavet}
\end{equation}
where $r_* \coloneqq r + 2M \ln (r/2M - 1) + R_*$ is the tortoise
coordinate, and $V$ is the effective potential. Here, for later
convenience, we intentionally keep an integration constant $R_*$ in the
tortoise coordinate defined by $\dv*{r_*}{r} = 1/(1-2M/r)$ for the
Schwarzschild spacetime. Unless otherwise stated, we set $R_* = 0$.

Throughout this article, we focus on the $l=2$ mode of axial
gravitational waves, for which the unperturbed effective potential is
given by the so-called Regge-Wheeler potential
\cite{1957PhRv..108.1063R}
\begin{equation}
 V_\mathrm{RW} (r_*) \coloneqq \pqty{1 - \frac{2M}{r}}
  \pqty{\frac{l(l+1)}{r^2} - \frac{6M}{r^3}} . \label{eq:rw}
\end{equation}
On another front, the polar gravitational waves obey
Eq.~\eqref{eq:wavet} with the so-called Zerilli potential
\cite{1970PhRvL..24..737Z}, which is known to be isospectral with the
Regge-Wheeler potential \cite{1975RSPSA.344..441C}. Vector and scalar
waves also obey the same equation with similar potentials. Thus, we
expect that the analysis of this study applies to fields of any parity
and spins, at least qualitatively.

The quasinormal-mode spectrum including the fundamental mode is reported
to be unstable when the potential is augmented by a bump modeled by the
P{\"o}schl-Teller form \cite{2022PhRvL.128k1103C},
\begin{align}
 V(r_*) & = V_\mathrm{RW} (r_*) + V_\mathrm{PT} (r_*; \epsilon , b) ,
 \label{eq:RWPT} \\
 V_\mathrm{PT} (r_*; \epsilon , b) & \coloneqq \frac{\epsilon}{(2M)^2
 \cosh^2[(r_* - b)/(2M)]]} , \label{eq:pt}
\end{align}
for sufficiently large values of $b$ and/or $\epsilon$.\footnote{Our
normalization is chosen to match that of
Ref.~\cite{2022PhRvL.128k1103C}, in which quantities are normalized by
$2M$. Our $b/M$ corresponds to their $2a$.} In this study, we fix
$\epsilon = \num{e-3}$, for which the fundamental mode is reported to be
destabilized if $b \gtrsim 50M$. We checked that the value of $\epsilon$
only determines the scale of the results presented in this study unless
it becomes $\sim \order{1}$.

\subsection{Time evolution} \label{sec:time_res}

To demonstrate that the ringdown signal is stable until the late-time
modes are excited, we numerically simulate scattering problems for
Eq.~\eqref{eq:wavet} with both the Regge-Wheeler potential of
Eq.~\eqref{eq:rw} and that perturbed by a P{\"o}schl-Teller bump of
Eq.~\eqref{eq:pt} adopting $b=100M$. The initial data are chosen to be
$\phi (t=0) = 0$ and the time derivative in the form of a Gaussian wave
packet,
\begin{equation}
 \pdv{\phi (t=0)}{t} = \exp \bqty{- \pqty{\frac{r_* - r_{*,0}}{2M}}^2} ,
\end{equation}
where $r_{*,0} = -50M$ or $50M$. Essentially the same initial data are
adopted in Ref.~\cite{2014PhRvD..89j4059B} for investigating the
scattering by double rectangular barriers (see also Sec.~\ref{sec:box}
below), and we have reproduced their results with reasonable
accuracy. The evolved fields are extracted at a finite distance of
$r_{*,\mathrm{ext}} = 250M$. The boundaries of our computational domains
are located at sufficiently large distances so that numerical
reflections do not affect the results shown below.

\begin{figure*}[tbp]
 \begin{tabular}{cc}
  \includegraphics[width=.48\linewidth,clip]{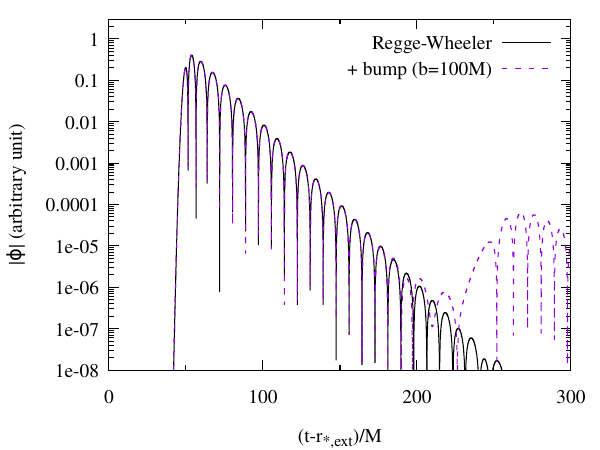} &
  \includegraphics[width=.48\linewidth,clip]{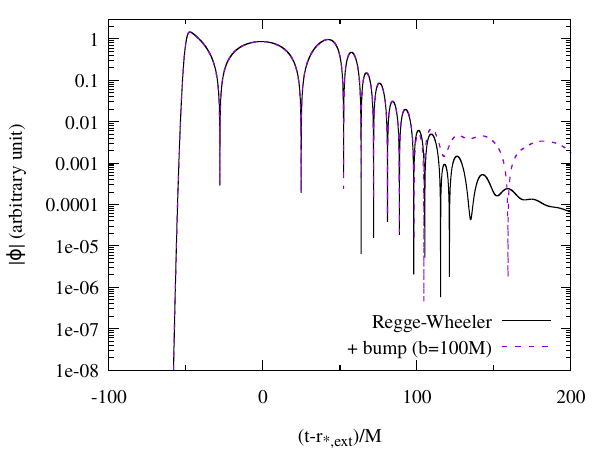}
  \end{tabular} \caption{Time evolution of a Gaussian wave packet in the
  Regge-Wheeler potential (black solid) and that perturbed by a
  P{\"o}schl-Teller bump (purple dashed). The P{\"o}schl-Teller bump is
  located at $b = 100M$ with the amplitude of $\epsilon = \num{e-3}$,
  with which the fundamental mode is destabilized
  \cite{2022PhRvL.128k1103C}. The Gaussian wave packets are initially
  located at $r_{*,0} = -50M$ and $r_{*,0} = 50M$ for the left and right
  panels, respectively. The origin of the time is chosen to account for
  the retardation at the extraction radius of $r_{*,\mathrm{ext}} =
  250M$. The power-law tail is also observed at $t - r_{*,\mathrm{ext}}
  \gtrsim 150M$ for the Regge-Wheeler potential with $r_{*,0} = 50M$
  \cite{1972PhRvD...5.2419P}.} \label{fig:timeevol}
\end{figure*}

Figure \ref{fig:timeevol} shows the time evolution of an initial
Gaussian wave packet in the Regge-Wheeler potential and that perturbed
by a P{\"o}schl-Teller bump. The left panel depicts the scattering of a
wave packet initiated at $r_{*,0} = -50M$, and this case corresponds to
outward propagation from the inside of the potential peak. It is obvious
that the ringdown signals, i.e., the time evolution of the field
characterized by exponential damping, are essentially unchanged under
perturbations until the echo signal arrives after $\sim 2b = 200M$ from
the direct signal. These features agree with those found in the analysis
of double rectangular barriers \cite{2014PhRvD..89j4059B} and the
Regge-Wheeler potential perturbed by a Gaussian bump
\cite{2022PhRvD.106h4011B}.

The right panel of Fig.~\ref{fig:timeevol} shows the scattering of a
wave packet initiated at $r_{*,0} = 50M$, which corresponds to the
middle of the main peak and the perturbative bump. This choice of the
initial location enhances the echo signal via the presence of
low-frequency components outside the main peak, which can be easily
reflected back and forth between the two potential barriers, from the
beginning \cite{2014PhRvD..89j4059B}.\footnote{In this study, the
``echo'' signal broadly denotes any late-time component resulting from
the existence of secondary potential peaks. The echo may be classified
into high-frequency components associated with repeated reflections of
unperturbed quasinormal modes and low-frequency components trapped in
the cavity between two potential peaks.} In the case of the right panel,
after the direct arrival of a precursor signal at $t-r_{*,\mathrm{ext}}
\sim -50M$, the ringdown signal sets in at $\sim 50M$ due to the
reflection from the Regge-Wheeler potential. Although the echo signal
begins to be dominant as early as $\sim 100M$ for the perturbed
spacetime, we can still observe the ringdown signal, whose oscillation
frequency and decay width are essentially unchanged from the unperturbed
values.

\subsection{Where have the quasinormal modes gone?}
\label{sec:time_freq}

Through the case studies shown in Fig.~\ref{fig:timeevol}, we may safely
conclude that the dominant part of the ringdown signal remains stable
despite destabilization of the quasinormal-mode spectra. This suggests
that \textit{bona fide} quasinormal modes defined as the poles of the
Green's function on the complex frequency plane may not always be useful
for characterizing observable signatures. A possible alternative may be
to rely on Wentzel-Kramers-Brillouin (WKB) and/or similar methods for
deriving approximate quasinormal-mode spectra
\cite{1985ApJ...291L..33S,1987PhRvD..35.3621I,1987PhRvD..35.3632I,2020PhRvD.101b4008H}. However,
this returns the same spectra as the unperturbed case in a trivial
manner, unless the potential is expanded to extremely high orders. We
would like to develop a method for approximately but stably inferring
unperturbed quasinormal modes, which are likely to characterize dominant
ringdown signals, and also for extracting influences of perturbations,
which are likely to introduce echo signals, simultaneously. Time-domain
analysis like Fig.~\ref{fig:timeevol} and that presented in
Refs.~\cite{2022PhRvL.128u1102J,2022PhRvD.106h4011B} is not fully
satisfactory, because the results may depend crucially on the initial
condition.

The key consideration is that the observable ringdown signals in the
time domain should be determined by the Green's function on the real
axis of the frequency, irrespective of whether they are evolved from
initial data or sourced by matter. In the remainder of this article, we
investigate scattering problems in the frequency domain. The Fourier
component of the field is governed by
\begin{equation}
 \dv[2]{\tilde{\phi}}{r_*} + [\omega^2 - V(r_*)] \tilde{\phi} = 0 .
\end{equation}
General solutions of this equation are asymptotically given by a
superposition of plane waves, $e^{\pm\mathrm{i} \omega r_*}$. We focus
on the ``in'' solution that satisfies purely downgoing boundary
condition at the horizon and hence behaves asymptotically as
\begin{equation}
 \tilde{\phi}_\mathrm{in} (r_*) =
  \begin{cases}
   e^{-\mathrm{i} \omega r_*} & (r_* \to -\infty) \\
   A_\mathrm{in} e^{-\mathrm{i} \omega r_*} + A_\mathrm{out}
   e^{+\mathrm{i} \omega r_*} & (r_* \to +\infty)
  \end{cases}
  , \label{eq:down}
\end{equation}
because it plays a central role in determining observable signals at
null infinity. The Green's function with downgoing (at the horizon) and
outgoing (at null infinity) boundary conditions is inversely
proportional to the Wronskian of corresponding homogeneous solutions,
which reduces to
\begin{equation}
 W = 2 \mathrm{i} \omega A_\mathrm{in}
\end{equation}
under the normalization of Eq.~\eqref{eq:down} and the similar one for
the ``up'' solution that satisfies purely outgoing boundary condition at
null infinity. Quasinormal modes are defined by zeroes of
$A_\mathrm{in}(\omega)$, which are equivalent to the poles of the
Green's function, and they never occur on the real axis due to the flux
conservation.

The fact that the zeroes of $A_\mathrm{in}(\omega)$ define the
quasinormal modes is reminiscent of the fact that the poles of the
$S$-matrix define the resonance in quantum systems (see, e.g., Sec.~13
of Ref.~\cite{taylor}). Let us recall the latter problem. For the
quantum scattering of a particle with mass $\mu$ by a
spherically-symmetric potential, the radial wave function $R_l (r)$ for
the $l$ mode is given by a spherical Bessel function. Its asymptotic
form is
\begin{equation}
 R_l (r \to \infty) \propto e^{-\mathrm{i} (kr - l\pi /2)} - S_l(E)
  e^{+\mathrm{i} (kr + l\pi /2)} , \label{eq:qmscat}
\end{equation}
where $E$ is the energy of the particle and $k^2 \coloneqq E/(2 \hbar^2
\mu)$. As depicted in the left panel of Fig.~\ref{fig:schematic}, the
first and second terms correspond to the incident and scattered waves,
respectively. If the scattering is elastic, the unitarity ensures that
the $S$-matrix has the unit amplitude, so that we may write $S_l (E) =
\exp[2\mathrm{i} \delta_l (E)]$ in terms of the (real) phase shift
$\delta_l (E)$. Meanwhile, the resonance is mathematically defined on
the analytically-continued complex energy plane as the pole $E =
E_\mathrm{res} \coloneqq E_\mathrm{R} - \mathrm{i} \Gamma_\mathrm{res}
/2$ of the $S$-matrix. Then, the $S$-matrix on the real axis is expected
to behave as
\begin{align}
 S_l (E) = e^{2\mathrm{i} \delta_l (E)} & \approx e^{2\mathrm{i}
 \delta_{\mathrm{BG},l} (E)} \frac{E - E_\mathrm{res}^*}{E -
 E_\mathrm{res}} \\
 & = e^{2\mathrm{i} \delta_{\mathrm{BG},l} (E)} \frac{E - E_\mathrm{R} -
 \mathrm{i} \Gamma_\mathrm{res} /2}{E - E_\mathrm{R} + \mathrm{i}
 \Gamma_\mathrm{res} /2} ,
\end{align}
where $\delta_{\mathrm{BG},l} (E)$ is the so-called background phase
shift and is expected to be a slowly-varying function of $E$. We stress
that the derivative of the phase shift behaves like a Lorentzian
function (see, e.g., Sec.~11.4 of Ref.~\cite{konishi_paffuti}),
\begin{equation}
 \dv{\delta_l}{E} \approx \frac{\Gamma_\mathrm{res}
  /2}{(E-E_\mathrm{R})^2 + (\Gamma_\mathrm{res} /2)^2} +
  \dv{\delta_{\mathrm{BG},l}}{E} .
\end{equation}
This quantity serves as a tool to extract the values of $E_\mathrm{R}$
and $\Gamma_\mathrm{res}$ via the location of a peak and the
full-width-half-maximum (FWHM), respectively, from the real-energy wave
function.

To apply this established tool in quantum mechanics to quasinormal modes
of black holes as a novel technique (see
Refs.~\cite{1994CQGra..11.2991A,1994CQGra..11.3003A,2003PhRvD..67l4017D,2009PhRvD..79j1501B,2021Univ....7..476H}
for related work), we assign ingoing (from null infinity) and downgoing
(to the horizon) waves, respectively, the roles of the incident and
scattered waves in quantum mechanics \cite{1990CQGra...7L..47G}. As
schematically presented in the right panel of Fig.~\ref{fig:schematic},
this identification allows us to investigate $A_\mathrm{in} (\omega)$ in
a similar manner to the Jost function used for defining the
$S$-matrix. Hereafter, we occasionally call $1/A_\mathrm{in} (\omega)$
the (complex) transmission amplitude.\footnote{Specifically in the
standard terminology of one-dimensional quantum scatterings, in which
$R$ and $T$ denote the reflection and transmission amplitudes for the
potential, respectively, $A_\mathrm{in}$ and $A_\mathrm{out}$ may be
expressed by $1/T$ and $R/T$, respectively.} Because $A_\mathrm{in}
(\omega)$ vanishes at the complex quasinormal-mode frequency, we expect
it to behave near the zero as
\begin{equation}
 A_\mathrm{in} (\omega) \approx \hat{A}_\mathrm{in} (\omega) \times
  (\omega - \omega_\mathrm{R} + \mathrm{i}\Gamma /2) ,
\end{equation}
where $\hat{A}_\mathrm{in} (\omega)$ is a slowly-varying and finite
function of $\omega$. Although the transmission amplitude is not unitary
and $\abs{A_\mathrm{in} (\omega)}$ is strictly larger than unity on the
real axis in our problem, this approximation derives a manifestly
unitary expression,
\begin{align}
 \frac{A_\mathrm{in} (\omega)^2}{\abs{A_\mathrm{in} (\omega)}^2} &
 \approx e^{-2\mathrm{i} \delta_\mathrm{BG} (\omega)} \frac{\omega -
 \omega_\mathrm{R} + \mathrm{i} \Gamma /2}{\omega - \omega_\mathrm{R} -
 \mathrm{i} \Gamma /2} , \\
 e^{-2\mathrm{i} \delta_\mathrm{BG} (\omega)} & \coloneqq
 \frac{\hat{A}_\mathrm{in} (\omega)}{\hat{A}_\mathrm{in}^* (\omega)} ,
\end{align}
on the real axis. Thus, by defining the phase shift of the transmission
amplitude as
\begin{equation}
 e^{-\mathrm{i} \delta (\omega)} = \frac{A_\mathrm{in}
  (\omega)}{\abs{A_\mathrm{in} (\omega)}} ,
\end{equation}
we arrive at the tool to extract quasinormal-mode frequency, i.e., the
derivative of the phase shift,
\begin{equation}
 \dv{\delta}{\omega} \approx \frac{\Gamma /2}{(\omega -
  \omega_\mathrm{R})^2 + (\Gamma /2)^2} +
  \dv{\delta_\mathrm{BG}}{\omega} .
\end{equation}
Indeed, this procedure is similar to the definition of the $S$-matrix
and the phase shift from the Jost function, while, unlike the case of
the quantum scattering, $A_\mathrm{out} (\omega)$ is not equivalent to
$A_\mathrm{in} (-\omega)$ and is not utilized here. If this phase shift
enables us to extract, at least approximately, the oscillation frequency
and the decay width of quasinormal modes for the unperturbed potential,
the same information may also be inferred from the phase shift for the
perturbed potential, taking the stability of ringdown signals into
account.

It should be remarked that our primary interest is in the functional
structure of $A_\mathrm{in} (\omega)$ such as zeroes and poles, and
whether the component is physically scattered or not is irrelevant
here. The $S$-matrix in quantum mechanics is referred only to explain
the idea behind our method. If we take the word ``scattering''
literally, the reflected, outgoing component is naturally regarded as
the scattered component. Actually, a lot of previous work on the
scattering and quasinormal modes for the black-hole spacetimes have been
performed based on this natural identification (see, e.g.,
Ref.~\cite{1988sfbh.book.....F} and references therein for reviews of
early work on the scattering problem). In this study, we neglect the
outgoing component, because the relevant Green's function is determined
entirely by $A_\mathrm{in} (\omega)$ of the ingoing component.

In fact, $A_\mathrm{out}$ has some drawbacks for defining the phase
shift. First of all, the phase of $A_\mathrm{out}$ depends on the choice
of the tortoise coordinate via $R_*$, while its magnitude for the real
frequency is simply given by $\abs{A_\mathrm{out}}^2 =
\abs{A_\mathrm{in}}^2 - 1$ via the flux conservation. Still, the
derivative of $\arg(A_\mathrm{out})$ may contain invariant information,
and actually, the derivative of the phase shift defined from the
scattering amplitude $A_\mathrm{out}/A_\mathrm{in}$ appears to work in a
similar manner to the one defined from $A_\mathrm{in}$ in
principle. However, from the technical point of view, the numerical
accuracy tends to be degraded at high frequencies, because
$\abs{A_\mathrm{out}}$ becomes small due to the very weak reflection by
the potential. Additionally, the weak reflection at high frequencies
makes $A_\mathrm{out}$ fairly sensitive to perturbations, so that its
phase changes rapidly in a manner not necessarily related to the
quasinormal modes. Hence, we focus on the phase shift defined from
$A_\mathrm{in}$ and its derivative in this study.

In the following, we apply the idea developed in this section to the
investigation of ringdown signals and their stability in the
Schwarzschild spacetime. However, we defer the discussion on the
Regge-Wheeler potential and its perturbation to
Sec.~\ref{sec:rw}. Instead, we start from investigating a toy model in
the next section.

\section{Toy model: rectangular barriers} \label{sec:box}

To identify the cause of quasinormal-mode instability and to test the
strategy presented in the previous section, it is instructive to study
an analytically-solvable toy model consisting of double rectangular
barriers. The potential is given by
\cite{2014PhRvD..89j4059B,2022PhRvL.128k1103C}
\begin{align}
 V(r_*) =
  \begin{cases}
   0 & (r_* < 0) \\
   V_0 & (0 \le r_* < a) \\
   0 & (a \le r_* < b) \\
   V_1 & (b \le r_* < b+d) \\
   0 & (b+d \le r_*)
  \end{cases}
  .
\end{align}
If $V_1 = 0$, this potential reduces to a single rectangular barrier
adopted by seminal work of Chandrasekhar and Detweiler
\cite{1975RSPSA.344..441C}, and the ``in'' solution in the form of
Eq.~\eqref{eq:down} satisfies
\begin{align}
 A_\mathrm{in} & = e^{+\mathrm{i} \omega a} \frac{2\omega k \cos (ka) -
 \mathrm{i} (k^2 + \omega^2) \sin (ka)}{2\omega k} , \\
 A_\mathrm{out} & = e^{-\mathrm{i} \omega a} \frac{\mathrm{i} (k^2 -
 \omega^2) \sin (ka)}{2\omega k} ,
\end{align}
where $k^2 \coloneqq \omega^2 - V_0$. If $V_1 \neq 0$, not necessarily
requiring $V_1 \ll V_0$, the solution changes to
\begin{widetext}
\begin{align}
 A_\mathrm{in} = \frac{1}{4\omega^2 k k'} \biggl( \{ & 4\omega^2 k k'
 \cos (ka) \cos (k' d) - (\omega^2 + k^2) (\omega^2 + k'^2) \sin (ka) \sin
 (k' d) \notag \\
 - & 2\mathrm{i} \omega [k (\omega^2 + k'^2) \cos (ka) \sin (k' d) + k'
 (\omega^2 + k^2) \sin (ka) \cos (k' d)] \} e^{+\mathrm{i} \omega (a+d)}
 \label{eq:ainbox} \notag \\
 + & (\omega^2 - k^2) (\omega^2 - k'^2) \sin (ka) \sin (k' d)
 e^{+\mathrm{i} \omega (2b-a+d)} \biggr) , \\
 A_\mathrm{out} = \frac{1}{4\omega^2 k k'} \{ [ & (\omega^2 + k'^2) \sin
 (k' d) - 2\mathrm{i} \omega k' \cos (k' d)] (\omega^2 - k^2) \sin (ka)
 e^{-\mathrm{i} \omega (a+d)} \notag \\
 - & (\omega^2 - k'^2) \sin (k' d) [(\omega^2 + k^2) \sin (ka) +
 2\mathrm{i} \omega k \cos (ka)] e^{-\mathrm{i} \omega (2b-a+d)} \} ,
\end{align}
\end{widetext}
where $k'^2 \coloneqq \omega^2 - V_1$.

\begin{figure*}[tbp]
 \begin{tabular}{cc}
  \includegraphics[width=.48\linewidth,clip]{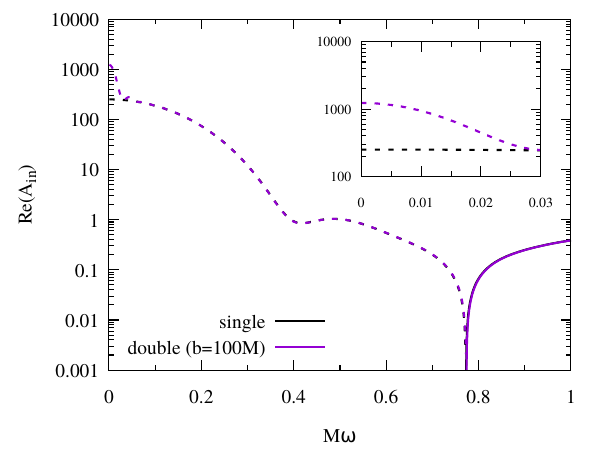} &
  \includegraphics[width=.48\linewidth,clip]{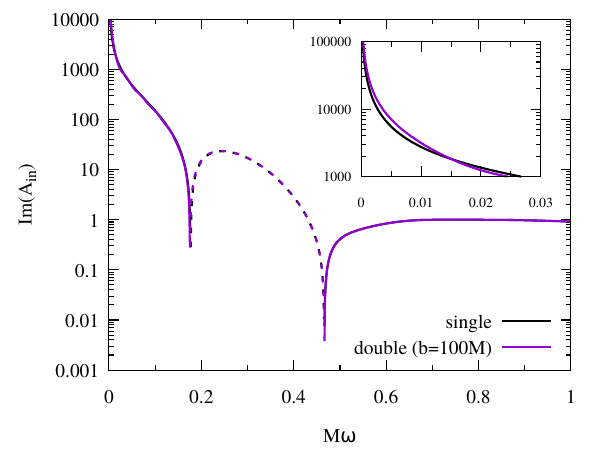}
 \end{tabular}
 \caption{Real (left) and imaginary (right) parts of $A_\mathrm{in}$ for
 the single and double rectangular barriers. The solid and dashed curves
 denote the positive and negative values, respectively. We set the
 parameters of the primary barrier to $V_0 = (0.4/M)^2$ and $a =
 9/(4MV_0)$ aiming at mimicking the height and the integral of the
 Regge-Wheeler potential. The second barrier is given by $b=100M$,
 $d=a$, and $V_1 = \num{e-3} /(2M)^2$. Curves in both panels are mostly
 indistinguishable for $M\omega \gtrsim \sqrt{V_1 M^2} = 0.016$ as
 depicted in the insets.} \label{fig:ainbox}
\end{figure*}

The real and imaginary parts of $A_\mathrm{in}$ for both the single and
double rectangular barriers are displayed in Fig.~\ref{fig:ainbox}. In
this and the following figures, we specifically adopt $V_0 = (0.4/M)^2$
and $V_0 a = 9/(4M)$ for the primary barrier to mimic the height and the
integral of the Regge-Wheeler potential. Parameters of the second
barrier is chosen to $b=100M$, $d=a$, and $V_1 = \num{e-3} /(2M)^2$. We
still normalize $V$ and $\omega$ by ``the mass of the black hole,'' $M$
as $V M^2$ and $M\omega$, although it is absent from the problem of
rectangular barriers. While the single barrier can be normalized by $a$
with leaving $V_0 a^2$ as the only parameter, introducing $M$ seems
equally useful for the case of multiple barriers.

Figure~\ref{fig:ainbox} indicates that $A_\mathrm{in}$ for the real
frequency does not change appreciably even in the presence of the second
barrier except for the low-frequency regime $M\omega \lesssim \sqrt{V_1
M^2} = 0.016$. This comparison strongly indicates the stability of the
ringdown signal. Indeed, while the perturbative second barrier also
destabilizes the quasinormal-mode spectrum \cite{2022PhRvL.128k1103C},
the dominant ringdown signal resulting from scattering is stably
characterized by the fundamental mode of the unperturbed potential
\cite{2014PhRvD..89j4059B}.

\subsection{Reason for the quasinormal-mode instability}
\label{sec:box_inst}

This toy model indicates that the quasinormal-mode instability is
relevant only for complex frequencies required by analytic continuation
of $A_\mathrm{in}$ given by Eq.~\eqref{eq:ainbox}, specifically the
factor $e^{2\mathrm{i} \omega b}$ appearing in the last line. On the one
hand, $A_\mathrm{in}$, the Wronskian, and the Green's function are
modified only on the order of $V_1 / V_0$ for $\omega \gg \sqrt{V_1}$,
as far as the frequency is real. Thus, the ringdown signals are also
kept unchanged within the order of perturbations, $V_1 / V_0$.

On the other hand, the negative imaginary part of the frequency, $\Gamma
> 0$, is the essential ingredient of the quasinormal modes. They
introduce a contribution in the form of $e^{\Gamma b}$, which grows
exponentially as the value of $b$ increases, i.e., the second barrier is
moved to a large distance. This exponential dependence is responsible
for the spectral instability of quasinormal modes on the complex
frequency plane (see also discussions in Ref.~\cite{1997PhRvL..78.2894L}
and Sec.~III C 1 of Ref.~\cite{2014PhRvD..89j4059B}). Actually, even for
the case of the Regge-Wheeler potential perturbed by a P{\"o}schl-Teller
bump, the threshold for the fundamental-mode instability shown in Fig.~2
of Ref.~\cite{2022PhRvL.128k1103C} appears to agree reasonably with the
dependence of $\epsilon e^{\Gamma b} = \mathrm{const.}$ expected for the
impact of the bump. To support this argument, we demonstrate that the
same mechanism works for generic two separated potentials in
App.~\ref{app:general}.

\subsection{Quasinormal mode from the phase shift} \label{sec:box_phase}

\begin{table}[tbp]
 \caption{Fundamental ($n=0$) and first two overtone ($n=1,2$) modes for
 the single rectangular barrier with $V_0 = (0.4/M)^2$ and $a =
 9/(4MV_0)$, which can be summarized as $V_0 a^2 = 2025/64$. The
 oscillation frequency and the decay width are presented by a normalized
 form of $M\omega_\mathrm{R}$ and $M\Gamma$, respectively. If the
 corresponding information can directly be extracted from the phase
 shift of the transmission amplitude $1/A_\mathrm{in}$, it is also
 presented as $M\omega_\mathrm{peak}$ and $M\Gamma_\mathrm{FWHM}$.}
 \begin{tabular}{ccccc}
  \hline
  $n$ & $M\omega_\mathrm{R}$ & $M\Gamma$ & $M\omega_\mathrm{peak}$ &
  $M\Gamma_\mathrm{FWHM}$ \\
  \hline
  $0$ & \num{0.4434} & \num{0.06319} & \num{0.4442} & \num{0.0566} \\
  $1$ & \num{0.5689} & \num{0.1918} & \num{0.5672} & N/A \\
  $2$ & \num{0.7458} & \num{0.3079} & N/A & N/A \\
  \hline
 \end{tabular}
 \label{table:qnmbox}
\end{table}

We move to extraction of quasinormal modes from the phase shift of the
transmission amplitude, $1/A_\mathrm{in}$, on the real axis of the
frequency. As a preparation, Table \ref{table:qnmbox} presents the
oscillation frequency $\omega_\mathrm{R}$ and the decay width $\Gamma$
for a single barrier. They are determined from the zeroes of
$A_\mathrm{in}$ in the complex frequency plane in a usual manner and are
also confirmed to characterize the ringdown signals. This table also
presents information extracted from the derivative of the phase shift,
$\omega_\mathrm{peak}$ and $\Gamma_\mathrm{FWHM}$, described below.

The derivative of the phase shift for a single rectangular barrier is
given by
\begin{equation}
 \dv{\delta_\mathrm{s}}{\omega} \coloneqq aV_0 \frac{ka [2\omega^2 -
  V_0 \sin^2 (ka)] - 2V_0 \sin (ka) \cos (ka)}{ka [4\omega^2 k^2 + V_0^2
  \sin^2 (ka)]} . \label{eq:phasebox}
\end{equation}
The full expression for the double rectangular barriers does not seem
very enlightening. Instead, we show the expression up to the linear
order of $V_1 / \omega^2$,
\begin{equation}
 \dv{\delta_\mathrm{d}}{\omega} \coloneqq \dv{\delta_\mathrm{s}}{\omega}
  \pqty{1 - \frac{V_1 A_1}{A_0}} + \frac{V_1 D}{A_0} ,
\end{equation}
where
\begin{widetext}
 \begin{align}
  A_0 & \coloneqq 4\omega^4 [4\omega^2 k^2 + V_0^2 \sin^2 (ka)] , \\
  A_1 & \coloneqq 4\omega^2 V_0 \{2\omega k \cos (ka) \cos [\omega (2b-2a+d)] -
  (\omega^2 + k^2) \sin (ka) \sin [\omega (2b-2a+d)]\} \sin (ka) \sin (\omega
  d) , \\
  D & \coloneqq D_o + D_a + D_d + D_b + 2\omega^2 d [4\omega^2 k^2 + V_0^2
  \sin^2 (ka)] , \\
  D_o & \coloneqq 4V_0 \Bqty{\frac{\omega^2 (\omega^2 + 3k^2)}{k} \cos (ka) \sin
  [\omega (2b-2a+d)] + \omega (3\omega^2 + k^2) \sin (ka) \cos [\omega
  (2b-2a+d)]} \sin (ka) \sin (\omega d) , \\
  D_a & \coloneqq - 4V_0 \omega^4 a \sin (\omega d) \sin [\omega (2b-2a+d)] , \\
  D_d & \coloneqq -2V_0 \omega^2 d \Bqty{2\omega k \cos (ka) \sin [\omega
  (2b-2a+d)] + (\omega^2 + k^2) \sin (ka) \cos [\omega (2b-2a+d)]} \sin (ka)
  \cos (\omega d) , \\
  D_b & \coloneqq 2V_0 \omega^2 (2b+d) \Bqty{(\omega^2 + k^2) \sin (ka) \sin
  [\omega (2b-2a+d)] - 2\omega k \cos (ka) \cos [\omega (2b-2a+d)]} \sin
  (ka) \sin (\omega d) .
 \end{align}
\end{widetext}
When the second barrier is located at a large distance, the term $D_b$
with an overall factor $2b+d \sim 2b$ introduces an apparently
nonperturbative correction to the derivative of the phase shift with the
period in frequency of $\Delta \omega = 2\pi /(2b-2a+d) \sim \pi /b$.

\begin{figure*}[tbp]
 \begin{tabular}{cc}
  \includegraphics[width=.48\linewidth,clip]{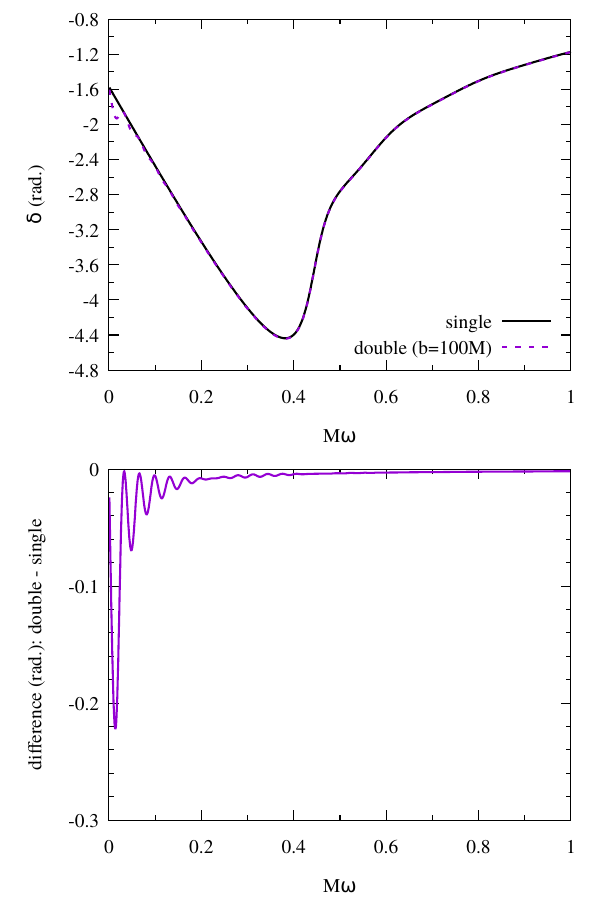} &
  \includegraphics[width=.48\linewidth,clip]{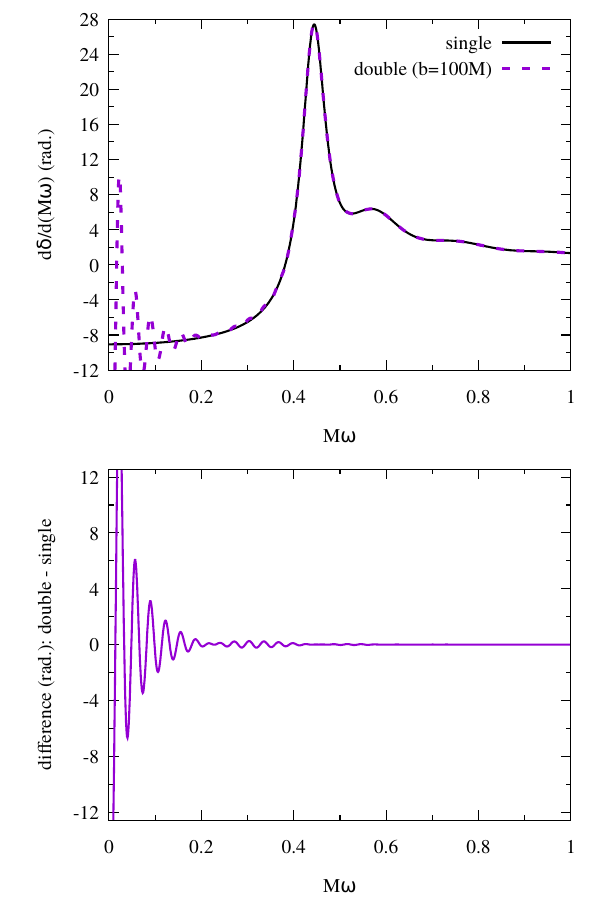}
 \end{tabular}
 \caption{Phase shift (left) and its derivative with respect to the
 frequency (right) for the single and double rectangular barriers (top)
 and their differences (bottom). The parameters of the potential are the
 same as those adopted in Fig.~\ref{fig:ainbox}. The primary peak of the
 derivative is located at $M\omega_\mathrm{peak} = 0.4442$ with the FWHM
 being $M\Gamma_\mathrm{FWHM} = 0.0566$. These values should be compared
 with $M\omega_\mathrm{R} = 0.4434$ and $M\Gamma = 0.06319$ for the
 fundamental mode of a single barrier (see Table
 \ref{table:qnmbox}). Another peak is also observed at
 $M\omega_\mathrm{peak} = 0.5672$, compared to $M\omega_\mathrm{R} =
 0.5689$ of the first overtone, although the FWHM cannot be extracted.}
 \label{fig:phasebox}
\end{figure*}

The phase shift and its derivative are shown in Fig.~\ref{fig:phasebox}
both for the single and double rectangular barriers. First, we focus on
the clean case of the single barrier. The derivative of the phase shift
clearly exhibits a sharp peak at $M\omega_\mathrm{peak} = 0.4442$ with
the FWHM being $M\Gamma_\mathrm{FWHM} = 0.0566$ (see Table
\ref{table:qnmbox}). The value of $\omega_\mathrm{peak}$ agrees with the
oscillation frequency $\omega_\mathrm{R}$ of the fundamental mode within
$\approx 0.2\%$. The value of $\Gamma_\mathrm{FWHM}$ agrees with the
decay width $\Gamma$ of the fundamental mode within $\approx 10\%$. This
analysis convinces us that the fundamental mode can approximately be
extracted from the phase shift defined from the real-frequency solutions
of the scattering problem for the case of a single rectangular barrier.

The deviation in the decay width may be ascribed, at least partially, to
contamination by a subdominant peak at $M\omega_\mathrm{peak} =
0.5672$. This value agrees with the oscillation frequency of the first
overtone, $M\omega_\mathrm{R} = 0.5689$, within $\approx 0.3\%$. Thus,
we believe that this subdominant peak is associated with the first
overtone, although the FWHM is no longer available in a straightforward
manner. We also see a shoulder-like structure around the oscillation
frequency of the second overtone, $M\omega_\mathrm{R} = 0.7458$, while
there is no extremum. Although we have not performed comprehensive
analysis, for a larger value of $V_0 a^2$, a larger number of peaks
associated with higher overtones become distinguishable. These
observations suggest that multiple quasinormal modes may be extracted
from $A_\mathrm{in}$ on the real axis by careful analysis. Indeed, this
must obviously be true if the higher overtones contribute to observable
ringdown signals
\cite{2019PhRvL.123k1102I,2021PhRvD.103l2002A,2022PhRvL.129k1102C,2022arXiv220202941I}. It
should be cautioned, however, that $\Gamma_\mathrm{FWHM}$ of the main
peak is not larger but smaller than $\Gamma$ of the fundamental mode for
the parameters adopted in this study. This implies that higher overtones
are unlikely to be extracted by simply assuming that
$\dv*{\delta}{\omega}$ is given by a sum of multiple peaks. Extraction
of multiple quasinormal modes is our ongoing project, and the result
will be presented elsewhere.

Remarkably, even if the second barrier is turned on and the
quasinormal-mode spectrum is destabilized \cite{2022PhRvL.128k1103C}, as
shown in Fig.~\ref{fig:phasebox}, the phase shift retains information
about the unperturbed quasinormal modes.\footnote{Strictly speaking,
while the overtones are destabilized, the fundamental mode is not
completely destabilized in our parameter set. We checked that our
argument still holds for a parameter set where the fundamental mode is
also destabilized.} The second barrier produces modulation in the phase
shift $\delta$ itself only on the order of $V_1 / V_0$ for the frequency
around the unperturbed quasinormal modes. While the modulation is
enhanced by a factor of $\sim 2b/M$ due to the differentiation, the
profile of $\dv*{\delta}{\omega}$ still recovers the unperturbed peaks
once averaged over the frequency range wider than $\Delta \omega \sim
\pi /b$.\footnote{We note that the value of $V_1 / V_0 \sim \num{e-3}$
adopted here is chosen to be so large that the effect of the second
barrier becomes visible on the scale of this figure.} This feature
explains why the initial stage of the ringdown signal in the perturbed
spacetime is characterized by the unperturbed quasinormal modes even if
the poles are destabilized on the complex frequency plane. Here, we
should recall that the fine resolution in the frequency domain becomes
relevant only when we have sufficiently long time-domain data. Thus, the
averaging over the frequency is naturally introduced when we focus on a
short time scale of $\lesssim 2b$ (see also
Refs.~\cite{2014PhRvD..89j4059B,2022PhRvL.128u1102J,2022PhRvD.106h4011B}
for relevant discussions).

For the case of double barriers, quasinormal modes associated with the
late-time echo signal may also be imprinted in the phase shift and its
derivative. The amplitude of the modulation in the phase shift grows as
the frequency decreases. Together with the enhancement by a factor of
$\sim 2b/M$, this dependence results in formation of equidistant narrow
peaks at low frequencies for the derivative of the phase shift. The
separation of $\Delta \omega \sim \pi /b$ between the maxima of these
peaks represents the travel time between the two barriers, and the
narrow widths may be consistent with long lifetimes. Thus, these peaks
may correspond to quasinormal modes associated with the echoes between
two barriers. However, because the peaks in $\dv*{\delta}{\omega}$
appear both as maxima and minima and its central value is displaced from
zero to $\approx -(4$--$5) \si{\radian}$, physical interpretations of
these peaks are not as straightforward as those of the main peak for the
unperturbed fundamental mode (see also Sec.~\ref{sec:rw}).

\begin{figure*}[htbp]
 \begin{tabular}{cc}
  \includegraphics[width=.48\linewidth,clip]{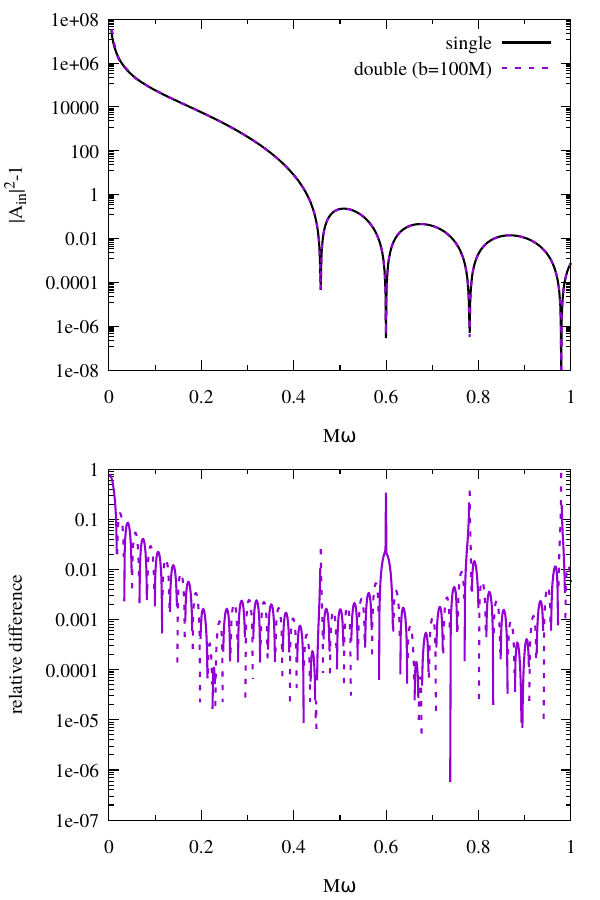} &
  \includegraphics[width=.48\linewidth,clip]{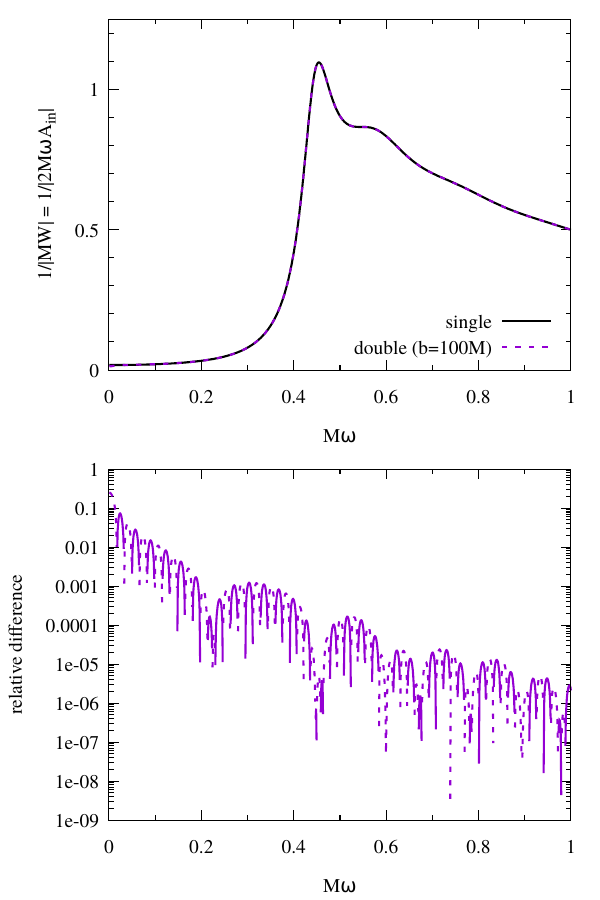}
 \end{tabular}
 \caption{Magnitude of the transmission amplitude in the form of
 $\abs{A_\mathrm{in}}^2 - 1 = \abs{A_\mathrm{out}}^2$ (left) and of the
 inverse of Wronskian, $1/\abs{W} = 1/\abs{2\omega A_\mathrm{in}}$
 (right) for the single and double rectangular barriers (top) and their
 relative differences (bottom). The parameters of the potential are the
 same as those adopted in Fig.~\ref{fig:ainbox}. The solid and dashed
 curves in the bottom panel represent positive and negative values,
 respectively, which indicate the excess and the deficit for the double
 barriers, respectively. The relative difference is enhanced for
 $\abs{A_\mathrm{in}}^2 - 1 \approx 0$ in the left panel.}
 \label{fig:magbox}
\end{figure*}

Before concluding this section, we mention that the magnitude of the
transmission amplitude on the real axis also reflects information about
the quasinormal modes, although not as clearly as the phase shift and
its derivative do. Figure~\ref{fig:magbox} shows the magnitude of
$A_\mathrm{in}$ in terms of $\abs{A_\mathrm{in}}^2 - 1 =
\abs{A_\mathrm{out}}^2$ and $1/\abs{W} = 1/\abs{2\omega A_\mathrm{in}}$
as functions of the real frequency. The oscillatory behavior of the
former is characteristic of the exactly rectangular potential, which
becomes transparent at discrete frequencies satisfying $\sin (ka) =
0$. The inverse of the Wronskian becomes large at these specific
frequencies as found in the right panel, and the overall features are
similar to $\dv*{\delta}{\omega}$ shown in the right panel of
Fig.~\ref{fig:phasebox}. Still, the peak frequency agrees only within
$\approx 5\%$ with the oscillation frequency of the fundamental mode,
compared to $\approx 0.2\%$ for the phase shift. Moreover, it is not
obvious how to extract decay widths from the broad profile. Thus, we
tentatively conclude that the phase is more useful than the magnitude
for extracting information about the ringdown signal. However, we recall
that both the magnitude and phase of $A_\mathrm{in}$ or $W$ on the real
axis are necessary to recover all the information on the complex
frequency plane.

\section{Phase shift for Schwarzschild black holes} \label{sec:rw}

\begin{figure*}[tbp]
 \begin{tabular}{cc}
  \includegraphics[width=.48\linewidth,clip]{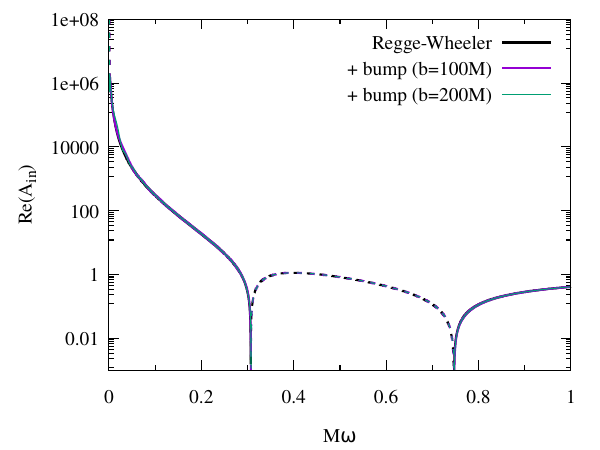} &
  \includegraphics[width=.48\linewidth,clip]{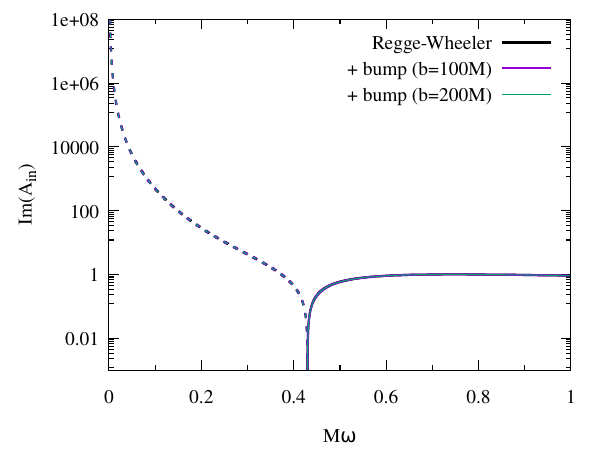}
 \end{tabular}
 \caption{Real (left) and imaginary (right) parts of $A_\mathrm{in}$ for
 the Regge-Wheeler potential (black) and that perturbed by a
 P{\"o}schl-Teller bump at $b=100M$ (purple) or $200M$ (green). The
 solid and dashed curves denote the positive and negative values,
 respectively. The magnitude of the bump is $\epsilon = \num{e-3}$. All
 the curves are mostly indistinguishable for $M\omega \gtrsim
 \sqrt{\epsilon} /2 = 0.016$.} \label{fig:ain}
\end{figure*}

We come back to our main problem of a Schwarzschild black hole and its
perturbation. Let us consider the potential given by
Eq.~\eqref{eq:RWPT}, i.e., the Regge-Wheeler potential augmented by a
small P{\"o}schl-Teller bump. Real-frequency ``in'' solutions in the
form of Eq.~\eqref{eq:down} are derived numerically for $b=100M$ and
$200M$, aiming at checking the dependence on $b$, as well as for the
unperturbed Regge-Wheeler potential. The real and imaginary parts of
$A_\mathrm{in}$ are shown in Fig.~\ref{fig:ain}. As we have noticed in
Sec.~\ref{sec:time}, a perturbative P{\"o}schl-Teller bump modifies
$A_\mathrm{in}$ only in a perturbative manner except for the low
frequency of $M\omega \lesssim \sqrt{\epsilon}/2 = 0.016$. In the
following, we present information about quasinormal modes contained in
$A_\mathrm{in}$ along the line presented in Sec.~\ref{sec:box_phase} for
the rectangular barriers. We remind that the fundamental mode of the
Regge-Wheeler (and also Zerilli) potential has the oscillation frequency
of $M\omega_\mathrm{R} = 0.3737$ and the decay width of $M\Gamma =
0.1779$ (see, e.g., Ref.~\cite{1985RSPSA.402..285L}).

\begin{figure*}[tbp]
 \begin{tabular}{cc}
  \includegraphics[width=.48\linewidth,clip]{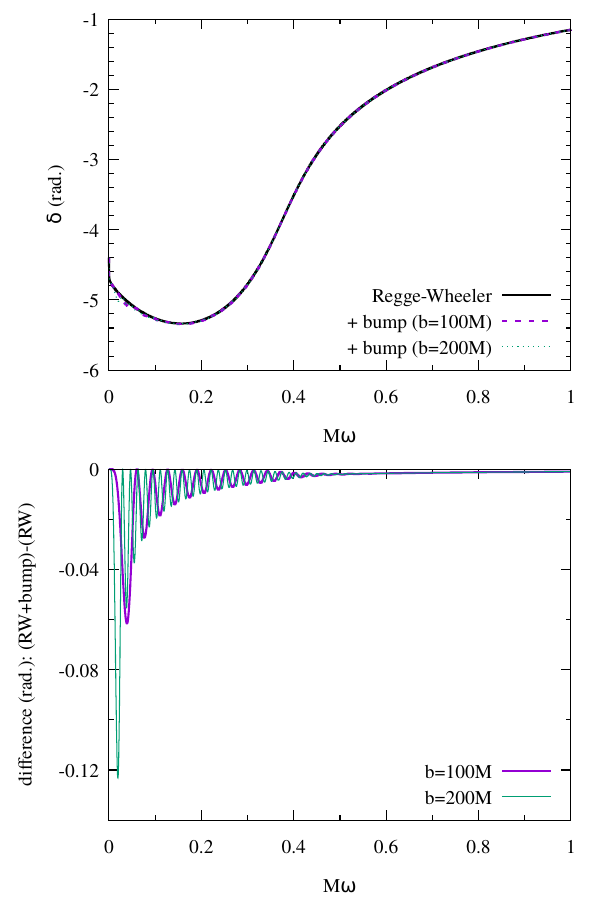} &
  \includegraphics[width=.48\linewidth,clip]{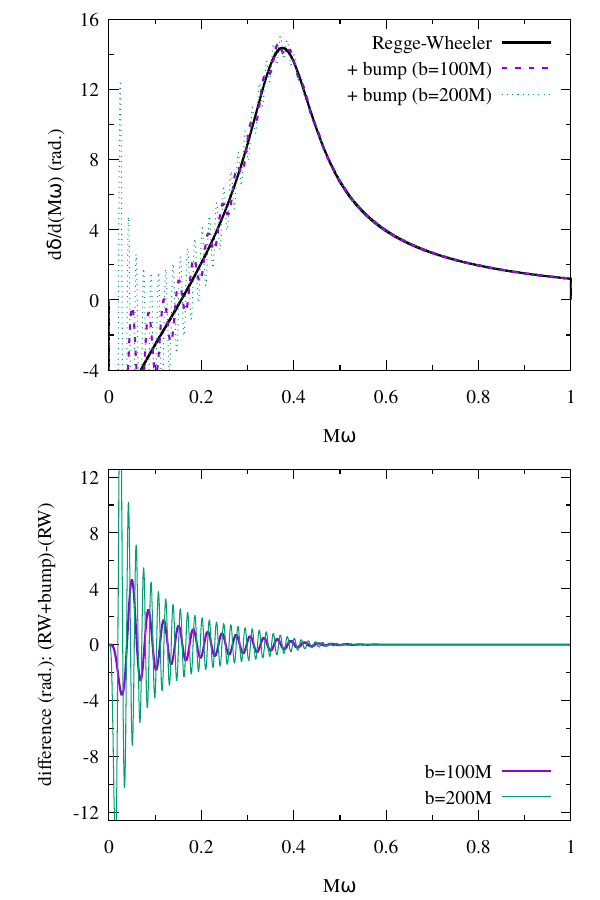}
 \end{tabular}
 \caption{Top: Phase shift (left) and its derivative with respect to the
 frequency (right) for the Regge-Wheeler potential (black solid) and
 that perturbed by a P{\"o}schl-Teller bump at $b=100M$ (purple dashed)
 or $200M$ (green dotted). The magnitude of the bump is $\epsilon =
 \num{e-3}$. For the Regge-Wheeler potential, the derivative peaks at
 $M\omega_\mathrm{peak} \approx 0.3758$ with the FWHM being $M
 \Gamma_\mathrm{FWHM} \approx 0.211$. These values should be compared
 with $M\omega_\mathrm{R} = 0.3737$ and $M\Gamma = 0.1779$ of the
 fundamental mode (see, e.g., Ref.~\cite{1985RSPSA.402..285L}). The
 modulation seen for the perturbed potential has a period of $\Delta
 \omega \approx \pi /b$, and its amplitude grows as the frequency
 decreases. Bottom: Differences between the quantities for the
 Regge-Wheeler potential and that perturbed by a P{\"o}schl-Teller
 bump.} \label{fig:phase}
\end{figure*}

Figure~\ref{fig:phase} shows the phase shift and its derivative with
respect to the frequency. Comparisons with the results for the single
rectangular barrier presented in Fig.~\ref{fig:phasebox} reveal that the
phase shift for the Regge-Wheeler potential exhibits a smooth structure
without a clear second peak. This is naturally understood by regarding
the Regge-Wheeler potential as a sequence of rectangular barriers with
various widths and heights (see also Ref.~\cite{1996PhRvD..53.4397N}).
Another way of explaining this smoothness is that the real parts of the
overtone modes are closer to the fundamental mode than in the case of
the single rectangular barrier. The single peak in
$\dv*{\delta}{\omega}$ of Fig.~\ref{fig:phase} is characterized by
$M\omega_\mathrm{peak} = 0.3758$ and $M\Gamma_\mathrm{FWHM} = 0.211$,
which agrees with the oscillation frequency and the decay width of the
fundamental mode of the Schwarzschild black hole within $\approx 0.6\%$
and $\approx 20\%$, respectively. The level of agreement is comparable
but worse by a factor of $2$--$3$ than the case of a single rectangular
barrier. This may again be ascribed to contamination by higher
overtones. Specifically, because the oscillation frequency of the first
overtone is lower only by $\approx 7\%$ than that of the fundamental
mode \cite{1985RSPSA.402..285L}, multiple poles are likely to be
contributing to the apparently single peak. This interpretation may be
supported by the fact that no isolated peak is found for higher
overtones.

The derivative of the phase shift allows us to understand the relation
between the quasinormal modes and the ringdown signal in an intuitive
manner. In quantum mechanics, the derivative of the phase shift has been
recognized to describe the time delay in the scattering process of a
wave with a given frequency \cite{1955PhRv...98..145W}. Because a wave
packet is formed at the frequency around which the time delay is nearly
stationary, the observable ringdown signal is naturally characterized by
the peak frequency of the derivative of the phase shift. This
interpretation appears consistent with the conventional understanding
that the quasinormal mode is the leakage of waves transiently trapped
around the light ring \cite{1972ApJ...172L..95G}. In turn, this
assessment suggests that the derivative of the phase shift is directly
related to observable ringdown signals even if the quasinormal-mode
spectra are destabilized.

The influence of the P{\"o}schl-Teller bump on the phase shift is only
perturbative, confirming the stability of the ringdown signal found in
Sec.~\ref{sec:time}. Because the modulation has a period in frequency of
$\Delta \omega \sim \pi/b$, the influence on the derivative of the phase
shift is enhanced by a factor of $\sim 2b/M$ (see the right panel of
Fig.~\ref{fig:phase}). Still, the overall shape of the peak is unchanged
around the unperturbed fundamental mode, and an appropriate averaging of
the phase shift (rather than its derivative) over the frequency will
allow us to extract the unperturbed fundamental mode in an approximate
but stable manner. Again, this feature should be the explanation for the
stability of dominant ringdown signals under the spectral instability of
quasinormal modes. At low frequencies, the amplitude of the modulation
grows in a similar manner to the case of double rectangular
barriers. This growth forms a bunch of peaks in the derivative of the
phase shift. These peaks are likely to represent quasinormal modes
associated with the echoes between the Regge-Wheeler potential and the
perturbative bump.

To sum up, the behavior of the phase shift for the Regge-Wheeler
potential with a possible perturbative bump is qualitatively the same as
that for the rectangular barriers. A notable difference is the absence
of distinguishable subdominant peaks associated with the higher
overtones. This may be reasonably explained by the proximity of the
poles for the unperturbed Regge-Wheeler potential.

\begin{figure*}[tbp]
 \begin{tabular}{cc}
  \includegraphics[width=.48\linewidth]{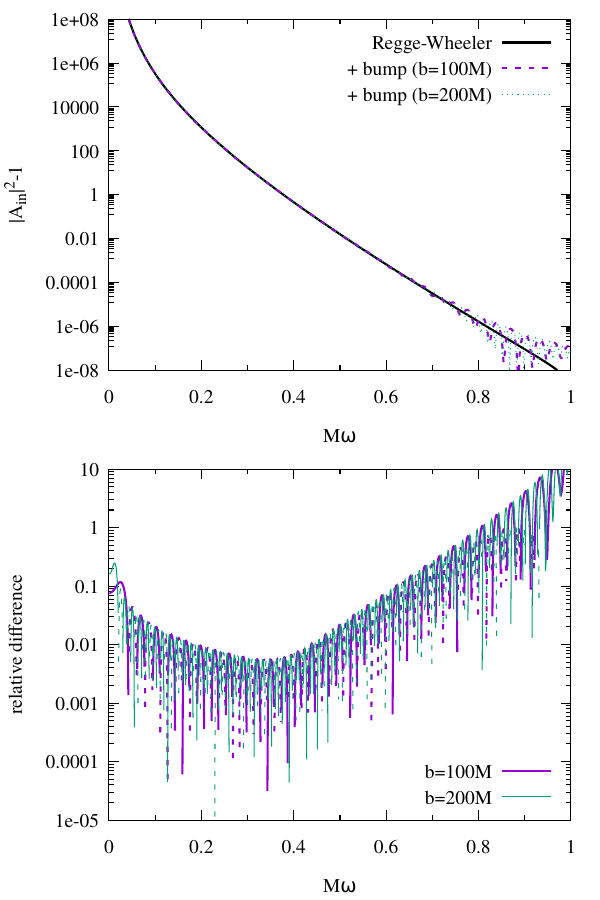} &
  \includegraphics[width=.48\linewidth]{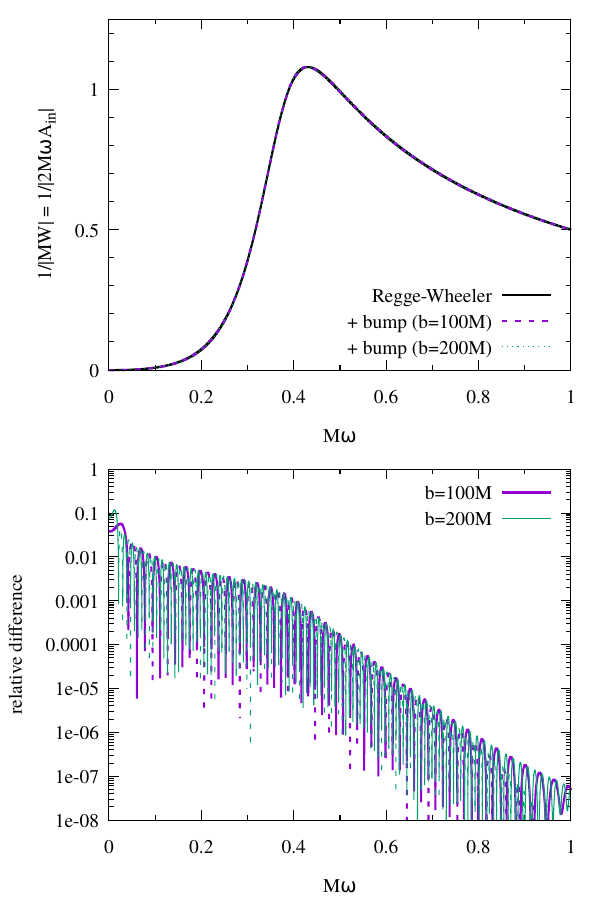}
 \end{tabular}
 \caption{Top: Magnitude of the transmission amplitude in the form of
 $\abs{A_\mathrm{in}}^2 - 1 = \abs{A_\mathrm{out}}^2$ (left) and of the
 inverse of Wronskian, $1/\abs{W} = 1/\abs{2\omega A_\mathrm{in}}$
 (right) for the Regge-Wheeler potential (black solid) and that
 perturbed by a P{\"o}schl-Teller bump at $b=100M$ (purple dashed) or
 $200M$ (green dotted). The magnitude of the bump is $\epsilon =
 \num{e-3}$. Bottom: Relative difference between the quantities for the
 Regge-Wheeler potential and that perturbed by a P{\"o}schl-Teller
 bump. The solid and dashed curves denote the positive and negative
 values, respectively, which indicate the excess and the deficit of
 values for perturbed potentials, respectively.} \label{fig:mag}
\end{figure*}

We also present information related to the magnitude of the transmission
amplitude in Fig.~\ref{fig:mag}. Differently from the case of
rectangular barriers, $\abs{A_\mathrm{in}}^2 - 1 =
\abs{A_\mathrm{out}}^2$ is a monotonic function except for tiny effects
of a perturbative bump. This is a mere repetition of the well-known fact
that the reflection and thus transmission coefficients of the
Regge-Wheeler potential are monotonic functions.\footnote{The
transmission coefficient $1/\abs{A_\mathrm{in}}^2$ is also called the
graybody factor in the context of black-hole evaporation
\cite{1974Natur.248...30H,1975CMaPh..43..199H,1976PhRvD..13..198P}.} The
magnitude of deviations observed for $M\omega \gtrsim 0.8$ is only on
the order of $\epsilon^2$, indicating that the reflection coefficient is
modified by filtering due to the P{\"o}schl-Teller bump on this
order. The differences in $\abs{A_\mathrm{in}}^2 - 1$ and $1/\abs{W}$
between the unperturbed and perturbed potentials change its frequency
dependence at $M\omega \approx 0.4$, which corresponds to the peak of
the Regge-Wheeler potential. The fact that the envelopes of these curves
are approximately independent of the value of $b$ is another indication
that the location of the perturbative bump does not influence the
stability of ringdown signals, although the quasinormal-mode spectrum
can be destabilized due to analytic continuation to the complex
plane. Instead, the dependence on $b$ resides in the period of the
oscillation, $\sim \pi /b$. While the inverse of the Wronskian seems to
contain some information about quasinormal modes in a similar manner to
the rectangular barriers shown in Fig.~\ref{fig:magbox}, it is not as
clear as the phase shift and its derivative do.

\section{Summary and discussion} \label{sec:summary}

We investigated the influence of proposed spectral instability of
quasinormal modes on the observable ringdown signal. We find that
perturbations to the potential is unlikely to destabilize the ringdown
signal, which is rather stably characterized by quasinormal modes in the
unperturbed spacetime except for the late-time echo signals. The
quasinormal-mode spectra owe their instability to analytic continuation
with respect to the frequency. Specifically, a phase factor of the form
$e^{2\mathrm{i} \omega b}$, with $b$ being the distance between the two
potentials, in $A_\mathrm{in}$ introduces an exponentially-growing
contribution once the frequency gains a negative imaginary part, which
in fact is the essential ingredient of the quasinormal modes.

Quasinormal modes of the unperturbed spacetime can be extracted from
real-frequency solutions of the scattering problem via the phase shift
defined from the transmission amplitude, $1/A_\mathrm{in}$ in our
notation, even if a perturbation destabilizes the quasinormal-mode
spectrum. This approach is motivated by the similarity of quasinormal
modes of black holes to resonances in quantum systems. Although this
extraction is not as precise as traditional methods such as the
continued fraction method \cite{1985RSPSA.402..285L}, the phase shift is
advantageous in its immunity to destabilization of \textit{bona fide}
quasinormal modes due to the analytic continuation. Rather, quasinormal
modes extracted from the phase shift may enable us to characterize
observable ringdown signals in an approximate but stable manner. The
implication of this method to the observable signal is that the dominant
ringdown signal is stably characterized by the unperturbed quasinormal
modes even under the spectral instability of quasinormal modes, and the
perturbative bump instead excites late-time echo signals, whose
amplitude is suppressed accordingly to the smallness of the bump. They
agree with and thus reconfirm the results of time evolution.

The analysis in terms of the phase shift may be applicable to a wide
range of spacetimes, most importantly the Kerr black hole and its
perturbed siblings. Potential difficulties, though not so serious, are
how to handle complex potentials associated with the black-hole spins
\cite{1973ApJ...185..635T,1982PThPh..67.1788S}, and transformation that
makes the potential real may be preferred
\cite{1976RSPSA.350..165C,1977RSPSA.352..381D}. It would also be
beneficial to perform systematic studies of quasinormal modes in
Schwarzschild spacetimes varying the parity, the spin of the field, and
the spherical harmonic eigenvalues.

After understanding that the ringdown signal is characterized by
information on the real axis, another question arises; why the ringdown
signal appears to be represented faithfully by the poles on the complex
plane for a simple potential like Regge-Wheeler's one? And, when the
quasinormal-mode spectrum is completely destabilized so that the pole of
the original fundamental mode disappears \cite{2022PhRvL.128k1103C}, how
do the remaining poles reproduce the original quasinormal modes in the
ringdown signal? It seems that a series of poles generated by
perturbations must be conspiring to preserve influence of the original
pole, at least of the fundamental mode, on the ringdown signal. This may
be accomplished by suitable distributions of the excitation factors
\cite{1986PhRvD..34..384L}. Our current study only considers the
frequency of quasinormal modes, and their excitation factor and/or
coefficient have not been investigated. Taking the fact that they should
also reside on the real axis into account, it is presumable that making
full use of $A_\mathrm{in}$ on the real axis should give us a clue to
these questions. Solving these problems will be helpful toward the era
of future detectors such as the Laser Interferometer Space Antenna
(LISA) \cite{2019arXiv190706482B}, with which ringdown signals will be
detected frequently with high signal-to-noise ratios, possibly
accompanied by noticeable influences of surrounding environments.

\begin{acknowledgments}
 We thank Kouichi Hagino, Hidetoshi Omiya and Norichika Sago for
 valuable discussions. This work was supported by Japan Society for the
 Promotion of Science (JSPS) Grants-in-Aid for Scientific Research
 (KAKENHI) Grant Numbers JP18H05236, JP20H00158, JP22K03617 (K.K.),
 JP18K13565, JP22K03639 (H.M.), and JP17H06357, JP17H06358, JP20K03928
 (T.T.).
\end{acknowledgments}

\appendix

\section{Spectral instability for generic two separated potentials} \label{app:general}

In Sec.~\ref{sec:box}, we demonstrated that the spectral instability of
quasinormal modes for double rectangular barriers is caused by a phase
factor $e^{2\mathrm{i} \omega b}$ by explicitly solving the scattering
problem. In this Appendix, we argue that this mechanism is a generic
feature of two separated scattering potentials relying only on the
asymptotic behavior of the solution (see also
Ref.~\cite{1997PhRvL..78.2894L}).

First, we consider two potentials $V_A (r_*)$ and $V_B (r_*)$, both of
which are localized around $r_* = 0$ and admit plane-wave solutions at
$r_* \to \pm \infty$. Specifically, the ``in'' solution that becomes
$e^{-\mathrm{i} \omega r_*}$ at $r_* \to -\infty$ may be expressed at
$r_* \to +\infty$ as
\begin{equation}
 \tilde{\phi}_\mathrm{in} (r_*) = A_\mathrm{in} e^{-\mathrm{i} \omega
  r_*} + A_\mathrm{out} e^{+\mathrm{i} \omega r_*} \label{eq:in_a}
\end{equation}
and
\begin{equation}
 \tilde{\phi}_\mathrm{in} (r_*) = B_\mathrm{in} e^{-\mathrm{i} \omega
  r_*} + B_\mathrm{out} e^{+\mathrm{i} \omega r_*}
\end{equation}
for the potentials $V_A (r_*)$ and $V_B (r_*)$, respectively. This also
implies that the solution that becomes $e^{+\mathrm{i} \omega r_*}$ at
$r_* \to -\infty$ is written as $B_\mathrm{in}^* e^{+\mathrm{i} \omega
r_*} + B_\mathrm{out}^* e^{-\mathrm{i} \omega r_*}$ at $r_* \to +\infty$
for the potential $V_B (r_*)$.

Next, let us consider the potential given by $V(r_*) = V_A (r_*) +
V_B(r_* - b)$, where $b$ is chosen so large that Eq.~\eqref{eq:in_a} is
approximately valid at $0 \ll r_* \ll b$. By translating the radial
coordinate, we readily found that the ``in'' solution for the potential
$V(r_*)$ takes the asymptotic form of
\begin{align}
 \tilde{\phi}_\mathrm{in} (r_*) & = (A_\mathrm{in} B_\mathrm{in} +
 A_\mathrm{out} B_\mathrm{out}^* e^{+2\mathrm{i} \omega b})
 e^{-\mathrm{i} \omega r_*} \notag \\
 & + (A_\mathrm{out} B_\mathrm{in}^* + A_\mathrm{in} B_\mathrm{out}
 e^{-2\mathrm{i} \omega b}) e^{+\mathrm{i} \omega r_*}
\end{align}
at $r_* \to +\infty$. Finally, the Green's function with downgoing and
outgoing boundary conditions is given in terms of the corresponding
Wronskian,
\begin{equation}
 W = 2\mathrm{i} \omega (A_\mathrm{in} B_\mathrm{in} + A_\mathrm{out}
  B_\mathrm{out}^* e^{+2\mathrm{i} \omega b}) , \label{eq:w_two}
\end{equation}
which involves the phase factor $e^{+2\mathrm{i} \omega b}$.

When the secondary potential $V_B (r_*)$ can be regarded as a
perturbation to the primary potential $V_A (r_*)$,
$\abs{B_\mathrm{out}/B_\mathrm{in}} \ll
\abs{A_\mathrm{out}/A_\mathrm{in}} \le 1$ holds. Thus, the second term
in the Wronskian, Eq.~\eqref{eq:w_two}, is perturbative for the real
frequency. However, the phase factor moves the poles of the Green's
function on the complex frequency plane in an outspiraling manner as the
value of $b$ increases
\cite{1997PhRvL..78.2894L,2014PhRvD..89j4059B,2022PhRvL.128k1103C},
causing the spectral instability.

%

\end{document}